\documentclass[conference]{IEEEtran}

\usepackage{cite}
\usepackage{amsmath,amssymb,amsfonts}
\usepackage{algorithmic}
\usepackage{graphicx}
\usepackage{textcomp}
\usepackage{xcolor}
\def\BibTeX{{\rm B\kern-.05em{\sc i\kern-.025em b}\kern-.08em
    T\kern-.1667em\lower.7ex\hbox{E}\kern-.125emX}}
\usepackage{orcidlink}
\usepackage{quantikz}
\usetikzlibrary{quantikz2}
\usepackage{adjustbox}

\newcommand{\Su}[0]{\mathrm{SU}}
\newcommand{\su}[0]{\mathfrak{su}}
\newcommand{\vt}[0]{\boldsymbol{\theta}}

\newtheorem{thm}{Theorem}

\newtheorem{proposition}[thm]{Proposition}

\newtheorem{defn}{Definition}

\begin{document}

\title{Subspace Controllability in Variational Quantum Circuits: Maximising Expressivity and Increasing Search Efficiency with Dynamical Lie Algebras\\
}

\author{
\IEEEauthorblockN{
Andrew Rowan Barlow\IEEEauthorrefmark{1}\IEEEauthorrefmark{2}\,\orcidlink{0009-0001-0883-7820}\,,
Hans-Martin Rieser\IEEEauthorrefmark{1}\,\orcidlink{0000-0002-1921-1436}\,, 
Markus Lange\IEEEauthorrefmark{1}\,\orcidlink{0000-0001-7198-2871}\,
}
\IEEEauthorblockA{
    \IEEEauthorrefmark{1}\textit{Institute for AI Safety and Security}, \textit{Deutsches Zentrum für Luft- und Raumfahrt}, Ulm/St. Augustin, Germany \\
    \IEEEauthorrefmark{2}\textit{Astronomisches Rechen-Institut}, \textit{Zentrum für Astronomie der Universität Heidelberg}, Heidelberg, Germany
}
}

\maketitle

\begin{abstract}
Variational hybrid quantum models are a common paradigm for realising machine learning on quantum hardware. Nonetheless, they are challenging to train due to several problems intrinsic to the loss landscapes formed by variational quantum circuits. To mitigate these issues, researchers have extended the idea of subspace controllability leading to an increase in the success rate of variational models in finding the global minimum for quantum-based tasks. However, it remains unknown whether these results extend to classical-based tasks, and if any advantages are realised over a hyperparameter search. We begin to address this question by employing controllable circuits in a multi-classification task using MNIST-1D. Our results show that over a hyperparameter search, the majority of quantum models that achieve the lowest training and validation losses are subspace controllable. These results indicate that the hyperparameter search can be restricted to such models, in turn reducing computational cost and time.
\end{abstract}

\section{Introduction}

The advancement of quantum computers presents new opportunities for exploring novel approaches for machine learning. A popular method is to train a parameterised quantum circuit by deferring the calculations for minimising the loss function to classical hardware; such models are referred to as variational hybrid quantum models. 

Training variational hybrid quantum models is a NP-hard task due to the nature of the loss landscape formed \cite{Bittel2021}. Some common difficulties that arise throughout training are exponentially vanishing gradients and barren plateaus which both hinder the model from finding the global minimum in the loss landscape \cite{McClean2018, Larocca2025, Fontana2024, Anschuetz2022}. It is further conjectured that the phenomenon of barren plateaus is intrinsic to any circuit that cannot be simulated efficiently on classical hardware \cite{Cerezo2025} using Lie algebraic simulation methods \cite{Goh2023}. 

Given the inescapable nature of this problem, a variety of techniques have been developed to mitigate these issues. These include systematically reducing the expressivity of the circuit to take advantage of symmetries within the dataset \cite{Nguyen2022, Meyer2022}, or employing an alternative definition of overparameterisation for quantum models that improves the success rate in learning quantum-based problems \cite{Larocca2023, Wiersema2023}. The mathematical framework that underlies most of this theory is the dynamical Lie algebra that is defined as the exhaustive commutation of Lie generators that generate each gate in the unitary circuit. This algebra offers a powerful tool in analysing any parameterised unitary circuit, and provides a way to study the expressivity of a circuit in terms of the unitary operators that it can express.

Borrowing the term from quantum optimal control theory, a circuit that can represent any operator in its operator Hilbert space is said to be subspace controllable, otherwise it is subspace uncontrollable. Subspace controllable circuits reduce the number of local minima as otherwise found in subspace uncontrollable circuits \cite{Russell2017, Wu2011}, and have been further demonstrated to increase the success rate in finding solutions to the transverse field Ising models and Heisenberg XXZ models \cite{AllenZhu2019}. The theory of quantum overparameterisation builds on subspace controllability for variational quantum circuits and has equally been successful in increasing the success rate in quantum-based learning tasks such as autoenconders, unitary compilation and Ising models \cite{Larocca2023}. 

Although empirical evidence suggest that subspace controllable circuits are advantageous for certain quantum-based learning tasks, where the data is quantum by nature, it is still unknown whether such circuits retain the same benefits when employed in \textit{classical}-based learning tasks, and furthermore in a practical machine learning setting, where a hyperparameter search is usually conducted.

This becomes an especially compelling question when noticing that the definition of subspace controllability is independent of the learning task, and so reported advantages found for quantum-based learning tasks should be transferable to classical-based tasks. Other questions are also raised, especially in the context of adding additional parameters to an already controllable circuit; the extra parameters are redundant in expressing any new operator. Therefore, how does the machine use the extra parameters in the quantum model? In the classical machine learning setting, this has non-trivial implications and leads to overfitting when near the interpolation threshold, and with further overparameterisation it may exhibit double-descent \cite{Belkin2019, Nakkiran2021}.

In this work, we investigate if subspace controllable circuits increase the success rate of learning a classical-based task of classifying three digits in MNIST-1D \cite{Greydanus2024}. Success is defined here by minimising the train and validation losses of models over a hyperparameter search, in order to make more relevant conclusions for employing variational hybrid quantum circuits in a practical machine learning scenario.

As typical with classical datasets and hyperparameter searches, a more complex picture emerges than reported in research with quantum-based tasks. Our experiments found that subspace controllable quantum models consistently achieve lower training losses than their subspace uncontrollable versions, even when the circuits had the same initial weight initialisation and hyperparameters. Although controllable circuits achieved the lowest losses, they also achieved the highest losses throughout their hyperparameter trials. A small set of hyperparameters caused this effect, which in turn made them perform worse than their uncontrollable counterparts. This runs contrary to the theoretical and practical results of previous works related to learning quantum-based tasks \cite{AllenZhu2019, Russell2017, Wu2011, Larocca2023}. The point at which the difference in the highest and lowest losses obtained throughout the hyperparameter search was minimised, was at the onset of the circuits becoming subspace controllable. 

Moreover, the majority of subspace controllable circuits that had additional parameters were observed to overfit their training data, in line with our theoretical results, that the additional parameters cannot increase the expressivity of the model and thus can for example be used by the machine learning algorithm for overfitting the data. Overfitting in quantum models has also been observed in \cite{Peters2023, Schuld2020}. In terms of overparameterisation, the saturation of expressivity when increasing the number of trainable parameters agrees with previous work \cite{AllenZhu2019, Larocca2023, Chen2021, Sim2019}, however the overall behaviour of circuits overfitting their data runs contrary to supporting evidence for the conjecture that unitaries provide regularisation \cite{Chen2021, Sim2019}, or experiments where minimal overfitting occurred \cite{Bowles2024}. Although select hyperparameter runs in our experiments would support these observations.

The greatest percentage of hyperparameter trials that minimised both training and validation losses contained circuits around their onset of controllability. This implies, for this ansatz and task, that an \textit{a priori} and clever choice of the number of trainable parameters increases the success rate for the corresponding machine learning models to learn a task over a hyperparameter search. This suggests that for future hyperparameter searches, the number of trainable parameters might be restricted, or even fixed, without affecting the learning process. This would subsequently decrease the computational time and overall expenses. 

Even though our results are limited to a single classical task, they warrant further research to apply the same methods when learning other classical-based tasks. 

The paper is organised as follows: Section \ref{sec:review} formulates the concept of circuit controllability and introduces the role of the dynamical Lie algebra. Section \ref{sec:experiment} overviews the learning task and experimental setup, with results analysed in Section \ref{sec:results}. Finally, the novelty and implications of this research is discussed along with potential future research directions in Section \ref{sec:conclusion}.

\section{Controllable circuits} \label{sec:review}

In order to understand what is required for a circuit to be controllable, the definition is constructed systematically through Lie algebra and their corresponding Lie groups applied in the framework of gate-based unitary circuits. It will become apparent that the question of controllability for the circuit ansatz of interest in our experiments simply rests on the dimension of the Lie algebra generated by the circuit. We also provide a general method for establishing if other circuit ansatzes are controllable. 

With the ability to construct and verify controllable circuits, the quantum model should have the maximum freedom to explore its loss landscape and find the optimal solution for a given task. In other words, the circuit is able to represent any unitary quantum circuit in its operator Hilbert space. Thus, if no satisfactory solution is found by the model, the expressivity of the model cannot be at fault.

\subsection{Expressing the operator space of circuits} \label{sec:lie-algebra}

The purpose in working with Lie algebra, both as a theoretical and practical tool, is the simplicity that it provides when analysing the parameters of quantum circuits. Sometimes it is easier to calculate and inspect the mechanics of the circuit in this Lie algebraic picture, rather than studying the unitaries themselves.

A unitary quantum circuit acting on $n$-qubits can be represented as a $2^n \times 2^n$ special unitary matrix. The special unitary Lie group, $\Su(2^n)$, is the space of all unitary matrices with positive unit determinant, and encompasses all possible unitary $n$-qubit quantum circuits up to a global phase. The associated Lie algebra is $\su(2^n)$, the space of all $2^n \times 2^n$ Hermitian matrices with zero trace.  
\begin{proposition}\label{prop:lie-connection}
    Any $N \times N$ Hermitian matrix with zero trace $T$, can be mapped to a $N \times N$ special unitary matrix $U$ via 
    \begin{equation}
        U = e^{iT}. \label{eq:suexp}
    \end{equation}
\end{proposition}
Note that the exponentiation is defined via the corresponding power series in $iT$. Proposition \ref{prop:lie-connection} is proven by showing that the definition of unitarity holds, i.e. $U^\dagger U = \mathbb{I_{N \times N}}$ is guaranteed using the exponentiated form in \eqref{eq:suexp} for all elements in $\su(N)$. The traceless requirement for elements in $\su(2^n)$ originates from the need of special unitary matrices to have unit determinant - which in turn removes the global phase that is  redundant upon measurement of the circuit.

Proposition \ref{prop:lie-connection} makes the connection between the special unitary Lie group and its Lie algebra for $n$-qubit circuits, and which is informally summarised by
\begin{equation}
    \Su(2^n) = e^{i \su(2^n)}. \label{eq:su-correspondence}
\end{equation}

Furthermore, the Lie algebra $\su(2^n)$ forms a vector space over $\mathbb{R}$, and so a basis can be constructed to represent any element in $\su(2^n)$.

The basis chosen for this paper is explicitly constructed using pauli strings, where the pauli matrices are
\begin{equation} \label{eq:paulis}
    X = \begin{pmatrix} 0 & 1 \\ 1 & 0 \end{pmatrix}, \,  Y = \begin{pmatrix} 0 & -i \\ i & 0 \end{pmatrix}, \, Z = \begin{pmatrix} 1 & 0 \\ 0 & -1 \end{pmatrix},
\end{equation}
and the notation of a $n$-qubit pauli matrix acting on the $j$th qubit where $j\leq n$ is
\begin{equation}
    P_j := \underbrace{\mathbb{I}_{2 \times 2} \otimes \cdots \otimes \mathbb{I}_{2 \times 2}}_{j-1}  \otimes P  \otimes \underbrace{\mathbb{I}_{2 \times 2} \otimes \cdots \otimes \mathbb{I}_{2 \times 2}}_{n-j}
\end{equation}
where $P \in \{X, Y, Z\}$, which is then used to build $n$-qubit pauli strings. An example of a 4-qubit pauli string is
\begin{equation}
    X_1Y_2Z_4 = X \otimes Y \otimes \mathbb{I}_{2\times 2} \otimes Z.
\end{equation}

\begin{proposition}\label{prop:su}
    A basis, $\mathcal{B}_n$, for $\su(2^n)= \textrm{span } \mathcal{B}_n$ is given by the concatenation of $n$-qubit pauli strings:
    \begin{multline}
        \mathcal{B}_n := \left\{ \sigma_1 \sigma_2 \cdots \sigma_n : \sigma_j \in \{\mathbb{I}_{2 \times 2}, X_j, Y_j, Z_j \} \right\}
     \\ \setminus \{ \mathbb{I}_{2^n \times 2^n} \}. \label{eq:pauli-basis}
    \end{multline}
\end{proposition}

This explicit construction of the basis provides the dimension of $\su(2^n)$. There are $4^n$ combinations of the pauli strings in \eqref{eq:pauli-basis}, and with the identity element removed, results in the vector space $\su(2^n)$ having dimension $4^n-1$. The subtraction of the identity accounts for the fact that the element $\mathbb{I}_{2^n\times 2^n}$ is related to the global phase and so must be removed for \textit{special} unitary groups. For a more operational argument, the final identity is not traceless and so is removed to form the complete set of Hermitian traceless matrices.

\begin{defn}
    Let $1 \leq m \leq 4^n - 1$. For any choice of nonzero parameters $\vt \in \mathbb{R}^m$ and generators $T_j \in \mathcal{B}_n$,
    \begin{equation}
        T_{\vt } := \sum_{j=1}^m \theta_j T_j  \in \su(2^n)  \end{equation}
    is a parameterised Hermitian matrix with $m$ parameters. 
\end{defn}

Defining the element $T_{\vt}$ in terms of a linear decomposition of Hermitian matrices introduces a way to constuct parameterised unitary matrices in $\Su(2^n)$ by using Proposition \ref{prop:lie-connection}.

Let $T_{\vt} \in \su(2^n)$ be a parameterised Hermitian matrix with $m$ parameters. Then, Proposition \ref{prop:lie-connection} induces the following map to express a parameterised unitary,
\begin{equation}
    U(\vt) = \exp{i \sum_{j=1}^m \theta_j T_j} \in \Su(2^n) \label{eq:unitary-hermitation-decomp}.
\end{equation}

In this sense, the matrices $T_j$ are often referred to as generators, as they generate a special unitary matrix. Example unitaries include the rotational X-gate, $R_X(\theta)$, that is commonly generated by 
\begin{equation}
    R_X(\theta) = e^{i \frac{\theta}{2} X},
\end{equation}
and another example involving pauli strings is the CNOT-gate with control qubit $k$ and target qubit $j$ \cite{Larocca2021}: 
\begin{equation}
    \textrm{CNOT}_{k \to j} = e^{i \frac{\pi}{4} (\mathbb{I}-Z_k - X_j + Z_k X_j)} \label{eq:cnot-example}
\end{equation}
where the parameter has been fixed to $\pi/4$.

Being able to build circuits with unitaries of the form found in \eqref{eq:unitary-hermitation-decomp} makes it interesting to study the image of these parameterised circuits, $\textrm{Im} \, U (\vt)$, over the domain of their parameters $\vt$. For instance, if $\textrm{Im} \, U  (\vt)= \Su(2^n)$, then $U  (\vt)$ can express every circuit with $n$ qubits. Otherwise, the image is restricted to a subset $M \subset \Su(2^n)$. 
If so, then $U$ can only express circuits in $M$.

\begin{defn}[Controllability]
A parameterised unitary circuit with $n$ qubits and $m$ parameters, $U: \mathbb{R}^m \to \Su(2^n)$, is \textit{controllable} on $M\subseteq \Su(2^n)$ when $U$ is surjective on $M$, i.e.,
\begin{equation}
    \forall V \in M \, \exists \vt \in \mathbb{R}^m :\, U(\vt) = V.
\end{equation}
\end{defn}

The controllability of parameterised circuits can often be established by using the following fact.
\begin{proposition} \label{prop:dim-alg}
    For any Lie algebra $\mathfrak{g}$ and covering Lie group $G$, the dimension of their vector spaces are equal,
    \begin{equation}
        \textrm{dim}\,  \mathfrak{g} = \textrm{dim} \,G.
    \end{equation} 
\end{proposition}

For instance, to determine if

\begin{equation}
    V(\theta_1, \theta_2, \theta_3) = e^{i \theta_1 X + i\theta_2 Y + i\theta_3 X} \label{eq:vexample}
\end{equation}

is controllable on $\Su(2)$, note that the generators only span a two dimensional space, which is explicitly the vector space $\textrm{span} \{ X, Y \}$. By Proposition \ref{prop:dim-alg}, $\textrm{Im} V(\vt)$ is two dimensional and so cannot be equal to $\Su(2)$ as it has 3 dimensions. Therefore, $V$ is uncontrollable on $\Su(2)$.

Ultimately, we wish to determine for some $n$-qubit \textit{gate-based} unitary circuit if it is controllable on some subset $M \subseteq \Su(2^n)$. If gate-based circuits were of the form found in \eqref{eq:unitary-hermitation-decomp} and \eqref{eq:vexample}, then this would be a trivial task as the controllability can be answered directly by finding the span of the generators in the exponent. However, quantum circuits are typically built via the concatenation of unitary gates generated by a single real scalar and generator, rather than a set of parameters and generators as found in \eqref{eq:unitary-hermitation-decomp}, i.e. a $n$-qubit gate-based parameterised circuit $U(\vt)$ with $m$ parameters is usually given as
\begin{equation}
     U(\vt)= \prod_{j=1}^m e^{i \theta_j T_j } \label{eq:circ-single-concat}
\end{equation}
where $T_j \in \mathcal{B}_n$ are chosen generators. 
Although the form of the circuit in \eqref{eq:circ-single-concat} makes the chosen parameters and generators explicit, it makes the determination of the controllability of the circuit on some operator space difficult, because it is not of the form given by \eqref{eq:unitary-hermitation-decomp}. 
Note however, that \eqref{eq:circ-single-concat} can be written in the form \eqref{eq:unitary-hermitation-decomp} (and vice versa) if all generators commute, i.e., $[T_i, T_j] = 0$ for all $i\ne j$.

Thus, in order to transform \eqref{eq:circ-single-concat} into a single exponential to determine the controllability on some subspace, the naive approach is to explicitly compute the generators resulting from the concatenation of the $m$ gates. This is evaluated via the Baker-Campbell-Hausdorf formula  \cite{Hall2003}, that expresses the product of two exponentiated matrices $A, B \in \su(N)$:
\begin{multline} \label{eq:bch}
    e^{iA} e^{iB} =  \exp \big\{ iA + iB - \frac{1}{2}[A,B] - i\frac{1}{12}[A, [A, B]] \\ - i\frac{1}{12}[B, [B, A]] + \mathcal{O} (A^2 B^3) + \mathcal{O} (A^3 B^2) \big\}
\end{multline}
where the commutation operation $[A, B] := AB - BA$ produces non-trivial results when $[A,B]\neq 0$, and the higher order terms are encapsulated in $\mathcal{O} (A^2 B^3)$ and $\mathcal{O} (A^3 B^2)$.

Note that the left hand side of \eqref{eq:bch} is the product of two special unitary matrices, thus the right hand side must also be a special unitary matrix. This is consistent with the fact that a Lie algebra is closed under the commutation operation by definition, and so the infinite series in the exponent of \eqref{eq:bch} can be resumed into an element of $\su(N)$.

Instead of explicitly calculating the product of exponentials in \eqref{eq:circ-single-concat} using the Baker-Campbell-Hausdorf formula, a more streamlined approach is to calculate on the level of the Lie algebra. In doing so introduces the exact set of generators required for studying controllability.
\begin{defn}[Dynamical Lie algebra]\label{def:dla}
    Let $U(\vt) = \prod_{j=1}^m \exp i  \theta_j T_j$ be a $n$-qubit gate-based circuit with $m$ independent parameters and generators $T_j \in \mathcal{B}_n$ for $1 \leq j \leq m$. Then, the \textit{dynamical Lie algebra} of $U$ is
    \begin{equation}\label{eq:dla}
        \textrm{dla}_U := \textrm{span} \langle \{ T_1, \cdots, T_m\} \rangle_\textrm{Lie}
    \end{equation}
    where $\langle \cdot \rangle_\textrm{Lie}$ denotes the exhaustive application of commutation operations of the elements within the set.
\end{defn}

Note how the operation $\langle \cdot \rangle_\textrm{Lie}$ is essentially calculating all the commutation elements that arise in the exponent of \eqref{eq:bch}.

Calculating the Dynamical Lie Algebra (DLA) of a circuit provides the dimension of the space spanned by the generators, and thus with Proposition \ref{prop:dim-alg}, allows us to deduce if the circuit is controllable on some given $M \subset \Su(2^n)$. However, when constructing a circuit with known generators, we are most interested in deducing if the constructed circuit is controllable for a certain subset of unitary space.
\begin{defn}[Dynamical Lie (covering) group] \label{defn:dlg}
    Let $U$ be a $n$-qubit circuit with $\textrm{dla}_U$ given in \eqref{eq:dla}. Then the \textit{dynamical Lie group} is 
    \begin{equation}
        \textrm{dlg}_U := \left\{ e^{\mathfrak{g}} \, \forall \, \mathfrak{g} \in \textrm{dla}_U \right\}.
    \end{equation}
\end{defn}

Determining if a circuit is controllable on $\textrm{dlg}_U$ is intuitively asking if there are enough parameters, and if they are distributed appropriately, to maximise the dimension of the operator space of the circuit. 
\begin{defn}[Controllable Circuit] \label{def:controllable}
    Let $U$ be as in Definition \ref{defn:dlg}. $U$ is said to be \textit{controllable} if $U$ is controllable on $\textrm{dlg}_U$.
\end{defn}

As an immediate consequence, if a circuit is controllable, then the number of parameters of the circuit must at least be the dimension of $\textrm{dlg}_U$, which is equal to the dimension $\textrm{dla}_U$. This follows from the fact that $\textrm{Im} \, U (\vt) = \textrm{dlg}_U$ by definition of controllability and Proposition \ref{prop:dim-alg}.

The process from beginning with a unitary gate-based circuit, and evaluating its controllability, is summarised in the following diagram which highlights the question that answers whether a circuit is controllable or not.
\begin{center}
\begin{equation} \label{eq:comm-diagram}
\adjustbox{scale=0.85}{
\begin{tikzcd}[wire types={n,n,n},  nodes={inner sep=4pt}, row sep=1.8em, column sep=3em]
    U(\vt) = \prod_{j=1}^m e^{i \theta_j T_j} \arrow[d] \arrow[dr, "\textrm{Im}"] &  &  \\
\{ T^j \} \arrow[d, "\langle \cdot \rangle_\textrm{Lie}"'] & \textrm{Im} \, U (\vt) \arrow[dr, dash, dashed, "\textrm{if equal as sets } \Rightarrow \textrm{ controllable}"] & \\
\textrm{dla}_U \arrow[rr, "\exp{ \cdot}"'] & & \textrm{dlg}_U
\end{tikzcd}
}
\end{equation}
\end{center}

So far, the full practicality of using Lie algebra has not been leveraged. The algebra has always been in the context of exponentiation of these circuits gates, but the real power in using this as both a theoretical and practical tool is to simply study the Lie algebra itself, instead of the parameterised unitary matrices. This is most practical when an ansatz to a circuit is supplied so that $\textrm{Im}\, U = \textrm{dlg}_U$ is guaranteed when the circuit contains a minimum number of independent parameters. With this setup, only the dimension of the DLA needs to be calculated. The explicit steps for doing so are:

\begin{enumerate}
    \item Obtain the set of generators that compose your gate-based unitary circuit, i.e. the generators found in the gates $e^{i\theta_j T_j}$.
    \item Calculate the DLA of the circuit by evaluating $\langle \{T_j\} \rangle_\textrm{Lie}$.
    \item Calculate the dimension of $\langle \{T_j\} \rangle_\textrm{Lie}$, which is simply the number of unique elements in the set generated by the Lie closure.
\end{enumerate}

In practice, and as performed in this paper for verification, the second step of calculating the set until exhaustion is computed by brute force. This can be costly, given the worse case upper bound scales exponentially with qubit size, i.e. $4^n-1$. A more computationally efficient approach is again to use a circuit ansatz that produces an already known DLA. 

\subsection{Controllability of the hardware efficient ansatz}

One of the most common ansatzes, and the one that is employed in the subsequent experiments for investigating the learning advantages of controllable circuits, is the Hardware Efficient Ansatz (HEA). This ansatz provides the full set of $\su(2^n)$ generators within a single layer \cite{Larocca2021}. As the DLA has been saturated within a single layer, the addition of extra layers does not increase the dimension of the DLA because its size is bounded, nor can the dimension decrease with additional unitary gate operations. Therefore any circuit with a single HEA layer produces a DLA that spans its full operator space, $\textrm{dlg}_U = \Su(2^n)$. 

Given the exponential dimension of the DLA in terms of qubit number, most HEA circuits in the NISQ era are uncontrollable. For instance, a 6 qubit HEA circuit with 3 parameters per qubit per layer requires a minimum number of 228 repeated layers in order to be controllable. This is beyond the number of layers for any meaningful NISQ era computation as it would be subject to extreme levels of noise on current hardware. 

This poses a problem even for simulating the quantum circuits on classical hardware, yet can be resolved by reducing the dimension of the circuit's DLA.

Certain ansatzes generate polynomial sized DLA, although their existence raises an important question: if there remains quantum advantage in using such ansatzes as techniques exist that can efficiently simulate such polynomial DLA circuits on classical hardware \cite{Goh2023}. So one can return to an exponentially scaling DLA to resolve the quantum advantage question, whereby another dilemma is revealed as it has been conjectured that an exponential DLA can induce barren plateaus \cite{Larocca2021, Fontana2024} in turn becoming detrimental for machine learning applications \cite{Cerezo2025}. This paper contributes to this area by providing an alternative: keep the exponentially scaling DLAs and provide the capability to mitigate difficulties in training by using controllable circuits.

It remains a broader question in how quantum models learn to utilise the extra parameters that do not increase the expressivity of the model. And if it's advantageous or detrimental to its learning, such as experiencing double descent or overfitting which are effects found in overparameterising classical models. This paper begins to address these questions for quantum models learning classical tasks.

\subsection{Data encoding determines the circuit architecture}

Studying the effects of controllable circuits for learning classical tasks requires an encoding to be created for embedding the classical data into the quantum circuit, which is otherwise not required for quantum-based learning tasks. An encoding for the classical data, $U^{(n)}_{\textrm{Enc}}$, is realised via a parameterised unitary,
\begin{equation}
    U^{(n)}_{\textrm{Enc}}: \mathbb{R}^{m} \to M \subseteq \Su(2^n)
\end{equation}
where $m$ is the number of features of the classical data and $M$ is a subset of the operator space. Then, the quantum state, $U^{(n)}_{\textrm{Enc}}(\boldsymbol{v})\ket{0}$, represents the vectorised classical data sample, $\boldsymbol{v}$. By viewing the encoding as another operator that embeds classical data, it becomes transparent that $\textrm{dla}_{U^{(n)}_{\textrm{Enc}}}$ will influence the overall circuit's DLA.

Therefore, reducing the DLA of a circuit for the task of learning \textit{classical} data necessarily implies that the encoding must be restricted to the subset of which the circuit is controllable on. 

For example, if a circuit is controllable on $\Su(2) \times \Su(2)$, then for guaranteeing that the controllability isn't lost, the encoding should produce an embedding of $\Su(2) \times \Su(2)$. Otherwise, if the encoding embeds data that spans $\Su(4)$, the whole circuit and its operator space subsequently expands to $\Su(4)$ and potentially makes the circuit non-controllable.

Ultimately the encoding introduces a shift in framing when constructing a parameterised circuit: the ansatz and the operator space it occupies must be based on the encoding.

In doing so, a balancing act must be performed where one would like to maximise the expressivity of $U^{(n)}_{\textrm{Enc}}$ while minimising the dimension of $\textrm{dla}_{U^{(n)}_{\textrm{Enc}}}$ for tractable computation. For our experiments, this balancing act resulted in an encoding that embeds data into $\Su(4) \times \Su(2)$.

\section{MNIST-1D Experiment} \label{sec:experiment}

The learning task chosen was an adapted form of the MNIST-1D dataset \cite{Greydanus2024}. Its creation was inspired from the handwritten digits of MNIST and intended to be an efficient and more distinguishable benchmark for deep neural networks. Our choice of this dataset was largely influenced by the benchmarking study in the work of \cite{Bowles2024}, that suggested this dataset for its low feature dimension and to further explore other domains where quantum machine learning could provide fruitful results. 

The authors compared the performance of logistic regression, multi-layer perceptron and convolutional models. As they did, we will use the first two models in our experiment as baselines, where the logistic regression is simply a single linear layer with a softmax function applied at the end. Due to the single layer and the 1-1 mapping, the input and output dimensions are fixed to the problem specifications. The multi-layer perceptron consists of three linear layers where the middle layer has an input and output dimension of 256. The first and second layers have a ReLu activation function applied after their output and a residual connection exists between the middle and last layers, where the residual weights are from the output of the first layer and ReLu activation function.

For our experiments, 5,000 samples of the digits 0, 2, and 6 were generated with the accompanying Python library of MNIST-1D. The data was created using the default arguments with minor modifications to restrict only the generation of the three classes. Each sample was normalised by dividing by the maximum absolute value across all 5,000 samples. This resulted in every feature value of each sample to lie within the range of $[- 1,1]$. The datasets were then finally shuffled. Throughout the training of all models, 10-fold cross validation was used on this set of digits.

Only classifying 3 numbers from MNIST-1D reduces the complexity and challenge of the classification task from the original. When restricted to these three digits, a logistic regression model obtains a macro f1-score in validation of 92\%. The classical multi-layer perceptron that appears in the MNIST-1D dataset will completely learn the task obtaining scores of over 99\%.

The difference between these scores is smaller than the differences of the validation accuracies obtained by the models when learning to classify the 10 digits. However, the low number of features needed for the classification of only 3 digits allows us to map these features to quantum circuits without further reducing the complexity of the data as for example necessary with methods such as principal component analysis that reduce the size of MNIST-1D images to a lower resolution. This loss of complexity would then raise the question if the quantum circuit is performing the bulk of the learning or if the preprocessing is already solving the task. Hence we decided to work with a learning task that has reduced learning hardness, in order to learn a dataset where the preprocessing retains the original form and complexities rather than removing difficult features such as the noise.

The preprocessing applied for this study preserves the original complexity of the data and does not provide any principal components for the quantum circuit to exploit. To verify that the features have not increased in linear separability, a logistic regression model was tested before and after the reprocessing was applied to the data. There was no difference in the validation scores of the best trials in the hyperparameter search, thus showing that the features of the data remain as linearly separable as they did without preprocessing.

For reasons that will be clarified later, the preprocessing amounts to taking the average of consecutive points for each image. Two classes of averaging were performed: one class averaged every ten features and the other class averaged every second feature, starting after removing the first and last two features from the image. These classes are named the \textit{macro} and \textit{meso} averages of the image. This preprocessing of the image results in a dimensional reduction from 40 to 20 features, where 16 features contain the meso information, and the final 4 contain the macro. This preprocessing is visualised in Fig. \ref{fig:pre-process-digits} for a selected set of digits.

The reason for processing the image data into meso and macro information is due to the circuit architecture which is designed to be controllable and tractable for simulation on classical hardware, whilst retaining enough expressivity to solve the learning task. 

\subsection{The controllable circuit}
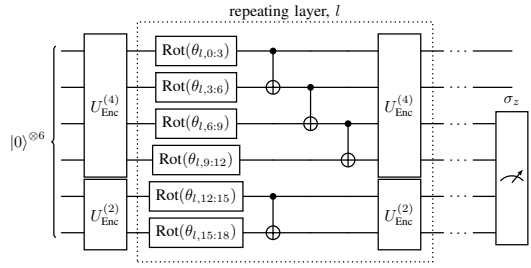
\begin{figure}[htbp]
\centering
\begin{tikzpicture}
\node[scale=0.6] {
\begin{quantikz}[row sep={0.8cm,between origins}]
    \lstick[6]{$\ket{0}^{\otimes 6}$} & \gate[4]{U_\textrm{Enc}^{(4)}} & \gate[1]{\textrm{Rot}(\theta_{l, 0:3})}\gategroup[6, steps=5, style={dotted}]{repeating layer, $l$} & \ctrl{1} & & & \gate[4]{U_\textrm{Enc}^{(4)}} & \ \ldots\ & \\
    & & \gate[1]{\textrm{Rot}(\theta_{l,3:6})} & \targ{} & \ctrl{1} &  &  & \ \ldots\ & \\
    & & \gate[1]{\textrm{Rot}(\theta_{l,6:9})} & & \targ{} & \ctrl{1} &  & \ \ldots\ & \meter[4]{\sigma_z} \\
    & & \gate[1]{\textrm{Rot}(\theta_{l,9:12})} & & & \targ{}  &  & \ \ldots\ &\\
    & \gate[2]{U_\textrm{Enc}^{(2)}} & \gate[1]{\textrm{Rot}(\theta_{l,12:15})} & \ctrl{1} & & & \gate[2]{U_\textrm{Enc}^{(2)}} & \ \ldots\ & \\
    & & \gate[1]{\textrm{Rot}(\theta_{l,15:18})} & \targ{} & & &  & \ \ldots \ & 
\end{quantikz}
};
\end{tikzpicture}\\
\caption{The circuit architecture chosen for experiments that becomes controllable when $l\geq15$, where $l$ denotes the number of repeating layers. The trainable parameters of the circuit are $\theta_{l,0\leq j < 18}$. The encoding map, $U^{(n)}_{\textrm{Enc}}(\boldsymbol{v})$, embeds a classical data sample $\boldsymbol{v} \in \mathbb{R}^m$ with $m$ features into a $n$-qubit state.}
\label{fig:quantum_circuit}
\end{figure}

The controllable circuit ansatz chosen is given in Fig. \ref{fig:quantum_circuit}. It is a modification of the usual HEA where the number of parameters is controlled via the number of re-uploading layers. The purpose of this architecture is for this circuit to become controllable after repeating 15 layers.

This architecture differs from a six qubit HEA by the removal of a CNOT-gate being applied on the 4th and 5th qubit, in effect creating two subsystems with 4 and 2 qubits. This decoupling allows for easier calculation of the circuit's DLA dimension, and further creates the opportunity to simulate these two systems in parallel, however this was not taken advantage of in the following experiments.

\begin{figure}[htbp]
\centerline{\includegraphics{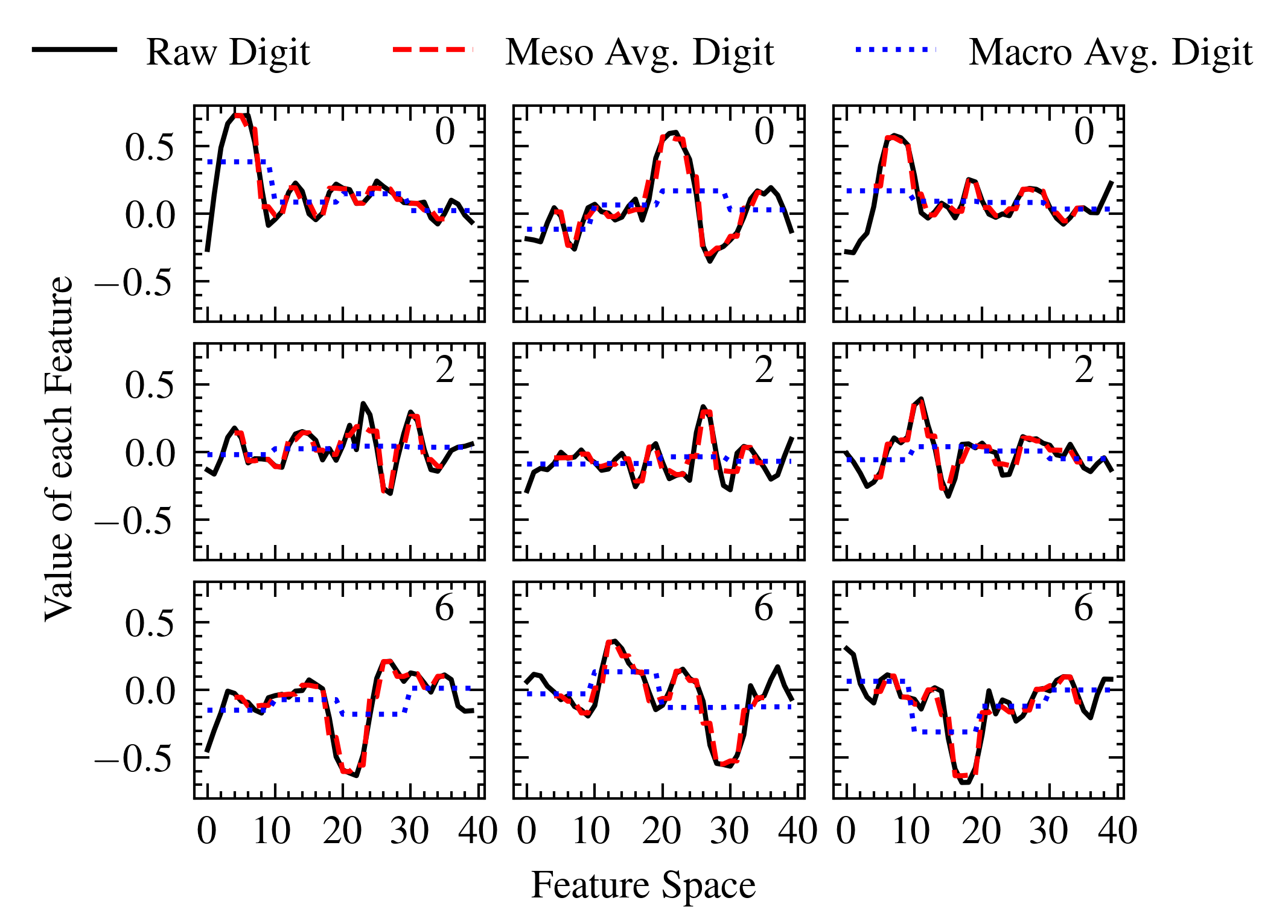}}
\caption{A visualisation of nine MNIST-1D digits, and their preprocessed versions `Meso Avg. Digit' and `Macro Avg. Digit', used in the multi-classification task. The number in the top right of each subplot denotes the label of the sample. The `Raw Digit' denotes the original digit, while the meso and macro digits are averaged over every two and ten consecutive features respectively.}
\label{fig:pre-process-digits}
\end{figure}

Following the discussions in Section \ref{sec:review}, the HEA ansatz guarantees that the circuit is controllable when there is an appropriate number of trainable parameters. As these two sub-systems are two decoupled HEA circuits, the DLA will not span the full $\Su(2^6)$ operator space but is calculated by considering all DLA generators from the 4 and 2-qubit subsystems embedded into the 6-qubit Hilbert space separately. Thus, the DLA is given by
\begin{multline}
    \langle \{T \otimes \mathbb{I}_{2^2\times2^2}\} \cup \{ \mathbb{I}_{2^4\times2^4} \otimes V \} \rangle_\textrm{Lie} \\
    = \langle \{T \otimes \mathbb{I}_{2^2\times2^2}\} \rangle_\textrm{Lie} \cup \langle \{ \mathbb{I}_{2^4\times2^4} \otimes V \} \rangle_\textrm{Lie} 
\end{multline}
for all $T \in \su(2^4)$ and $V \in \su(2^2)$.
Therefore, the dimensions of each subsystems' DLA are simply added to find the dimension of the combined system's DLA. Given that the 6-qubit circuit scales with $18l$ parameters with the number of layers $l$ as shown in Fig. \ref{fig:quantum_circuit}, then the onset of controllability is at layer $l \geq  (4^4+4^2-2)/18 = 15$. Hence, our circuit ansatz is controllable when $l \geq 15$.

For the DLA dimension to be conserved at encoding, it must also follow the 4 and 2-qubit decoupling architecture. The number of features that a logarithmic scaling encoding uses is derived from the summation of the dimension of the Hilbert space of each subsystem, $2^4+2^2=20$. This is exactly the number of features a preprocessed MNIST-1D digit contains, and was the original impetus for the choice in preprocessing.

The mapping of the preprocessed data onto an $n$-qubit circuit is given by the following logarithmic scaling encoding: 
\begin{align}\label{eq:encoding}
    & U_\textrm{Enc}^{(n)}: \, \mathbb{R}^{2^n} \to \Su(2^n) 
  ,\quad \boldsymbol{v} \mapsto H^{\otimes n} \bigoplus^{2^n}_{j=1} e^{i v_j} \, \mathbb{I}_{1\times1}
\end{align}
where $v_j$ are features of a data sample $\boldsymbol{v} \in \mathbb{R}^m$ assuming that $m\leq 2^n$ and $H$ is a Hadamard gate. The implementation of such an encoding using a universal set of gates scales logarithmic in circuit depth and approximately linear in spacetime with respect to the number of qubits, but these can be reduced if one allows an error in the mapping of the eigenvalues \cite{Gui2024}. Some addition in error while maintaining generalisation accuracy may be possible given the fact that MNIST-1D already contains noise in its features, however this question is deferred to future research.

The chosen encoding exhibits the desirable property of label invariance under a global translational shift in the y-axis of the data in Fig. \ref{fig:pre-process-digits}, as any global shift in the y-axis direction results in a global phase that is made redundant upon measurement.

After the preprocessed data is encoded into the qubits and ran through the circuit, the last 4 qubits are measured in the Z basis which results in the two subsystems comprising of meso and macro information of the digits. These 4 measurements are mapped using a fully connected layer with bias to a 3 dimensional output. In doing so, the importance of the meso and macro information is learnt through a linear weighting. Lastly, a softmax was added to the final 3 outputs to produce a probability distribution to classify the 3 digits.

Even though a classical layer exists and aids the quantum model in learning, it cannot be solely relied upon to solve the task as it only has 16 trainable parameters. As mentioned previously, the logistic regression only obtains 92\% validation score and has $40\times3=120$ trainable parameters. The quantum models obtain higher validation scores, and so means that the quantum model must be contributing to the learning of the task in a meaningful way.

For the learning, a hybrid approach was taken where the updates of the weights on the quantum circuit were performed classically. The quantum circuits were simulated without shots nor noise. The loss function selected was the cross entropy along with the Adam optimiser for minimising the loss function.

\subsection{Hyperparameter search}

To compare controllable and uncontrollable quantum circuits, a set of circuits were simulated where each individual set had a fixed number of repeated layers ranging from 7 to 25. For each set a hyperparameter study ran with 100 trails that were sampled randomly, where the hyperparameters and their respective ranges are defined in Table \ref{tab:hyp} for all models. The random search accounted for a subset spanning $8.\dot{8}\%$ of the total parameter space. Across each set, the sampled hyperparameters, and initialisation of circuit weights, remained the same - allowing for sole comparison of the sets that only differed in the number of trainable parameters. Every model had its weights initialised with a Gaussian distribution centred at 0 with standard deviation $\pi$ and $0.01$ for the quantum and classical models respectively. They were all trained for 100 epochs.

\begin{table}[htbp]
\caption{Hyperparameter of Models}
\begin{center}
\begin{tabular}{lr}
\textbf{Hyperparameter}& \textbf{Set of Possible Values} \\
\hline
learning rate, $\alpha$ & 0.001, 0.01, 0.1 \\
$\beta_1$ & 0.85, 0.875, 0.9, 0.925, 0.95 \\
$\beta_2$ & 0.95, 0.96, 0.97, 0.98, 0.99 \\
batch size & 25, 50, 100 \\
scaling parameter$^{\mathrm{a}}$ & $\frac{\pi}{4}, \frac{5\pi}{16}, \frac{3\pi}{8},\frac{7\pi}{16}, \frac{\pi}{2} $ \\
\hline
\multicolumn{2}{l}{
$^{\mathrm{a}}$Only quantum models included this hyperparameter.
}
\end{tabular}
\end{center}
\label{tab:hyp}
\end{table}

The same training procedure was conducted for the HEA and classical models, where the classical models only ran for a single hyperparameter search as they do not contain repeating layers.

Once the training for all models were finished, the top 5 best performing models from each hyperparameter trial were selected to train for another 250 epochs on a new set of 1,000 data samples with a 80/20 train and validation split. These models established the generalisation accuracies and thus ensured that the best performing hyperparameters were not fitted to their initial training set.

\subsection{Libraries}

The models were built using Flax 0.10.6 \cite{Heek2024}, a machine learning framework that is based on Jax 0.7.0 \cite{Bradbury2018} along with the Optax 0.2.5 \cite{DeepMind2020} to optimise the models. The hyperparameter search was conducted with Optuna 4.4.0 \cite{Akiba2019} and model tracking with MlFlow 3.1.4 \cite{Chen2020}. The configuration of the studies were aided with Hydra 1.3.2 \cite{Yadan2019}. The quantum simulations were performed using Pennylane 0.42.1 \cite{Bergholm2018}.

\section{Results} \label{sec:results}

\subsection{Test performance}

The logistic regression models had the lowest test f1-scores out of all the ansatzes, where the top 5 runs had achieved just under $92\%$. The quantum circuits, both HEA and controllable ansatzes, had achieved a similar score when the number of repeating layers were 7. It's noteworthy that the number of parameters in the logistic regression models is near the number of trainable parameters in 7 repeated layers of the quantum circuits.

The multi-layer perceptron outperformed all models in convergence speed, generalisation accuracies and the number of successful models throughout hyperparameter searches. As the classical model was trained on the preprocessed data, this proves that the preprocessing still allows the possibility for perfect classification. Neither did the logistic regression nor quantum models achieve perfect validation scores of over $99\%$.

The differences between the performance of the HEA and controllable ansatz in training and generalisation accuracies were minimal. Although the HEA is uncontrollable throughout all tests, it consistently achieved similar losses and generalisation accuracies throughout the hyperparameter searches. The only similarity in each architecture was the encoding, which lies in the space of $\Su(2^4) \times \Su(2^2) \subset \Su (2^6) $. The similar behaviour in training and generalisation indicates that the encoding's operator space has a large influence in the parameter set that optimises the circuit. This doesn't conclude that the HEA finds the same optimum as the controllable ansatz, yet it does suggest that the uncontrollable HEA is not disadvantaged in finding solutions for an encoding embedded in a smaller space, when compared to an adapted ansatz with an operator space wholly embedded in the space of the encoding. 

The observed results were that the HEA ansatz behaved as if it was restricted to the controllable subspace, and thus showed similar behaviours. Of course, simulating the full HEA ansatz is more costly than the controllable ansatz in qubit space, and so the controllable ansatz remains most practical. Throughout the remaining discussion, the HEA is considered to have achieved the same results as the controllable ansatz.

For the controllable ansatz, the highest test accuracies for quantum models ranged between $95\%$ and $96\%$. The highest achieving models were consistently found to be in their controllable regime, where the highest test accuracies of uncontrollable circuits were the least performant. This is comparative in performance to previous quantum circuit benchmarking studies that achieved around 96\% validation accuracy when classifying only digits 3 and 5 after applying PCA \cite{Bowles2024, AlvarezEstevez2025}.

\begin{figure*}[htbp]
\centerline{\includegraphics[width=\textwidth]{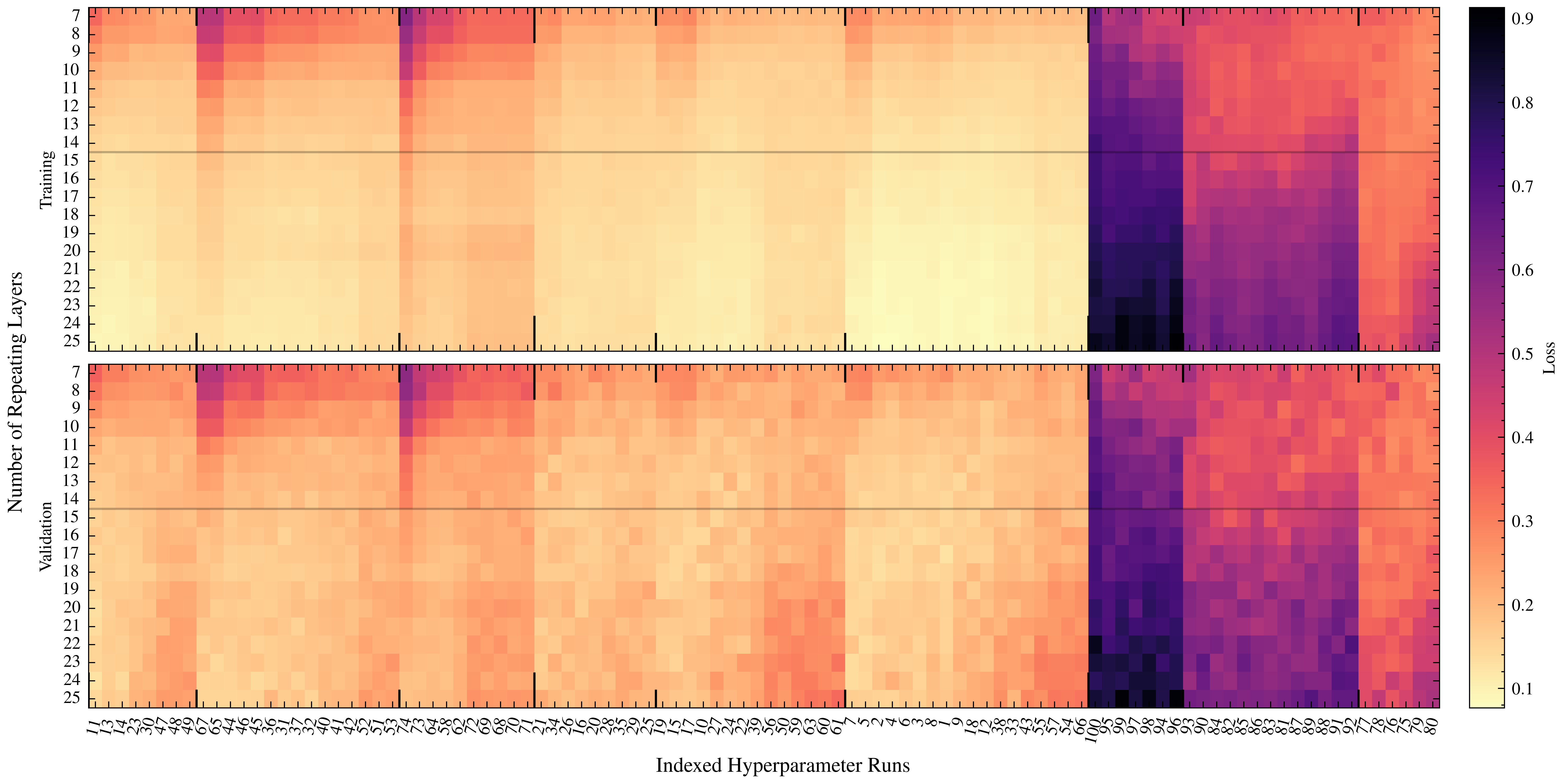}}
\caption{
    An overview of the performance of controllable circuits over all hyperparameter trials conducted in the experiment. Each cell's value is the aggregated final loss of the 10 models that were trained for each hyperparameter sample. The aggregation method used was the inter-quartile mean. Each column represents a fixed hyperparameter configuration and is sorted ascending from left to right based on the learning rate of the Adam optimiser, then batch size, and finally scaling parameter. The large and medium ticks represent a change in learning rate and scaling parameter respectively. The index of the hyperparameter sample along the $x$-axis is the rank in minimising the sum of the validation losses across the column. The  index on the $y$-axis are the number of repeating layers in the circuits across the row, where each layer contains 18 trainable parameters. The circuits become controllable when they have 15 or greater number of repeated layers, represented by the horizontal line.  Top \& Bottom: the final loss values of the models throughout their 100 training and validation epochs respectively. This figure demonstrates that circuits which minimised their training and validation losses over a hyperparameter search were controllable, where a minority of runs with the largest learning rate performed worse when controllable.
}
\label{fig:hyp-heatmap}
\end{figure*}

\begin{figure*}[htbp]
\centerline{\includegraphics{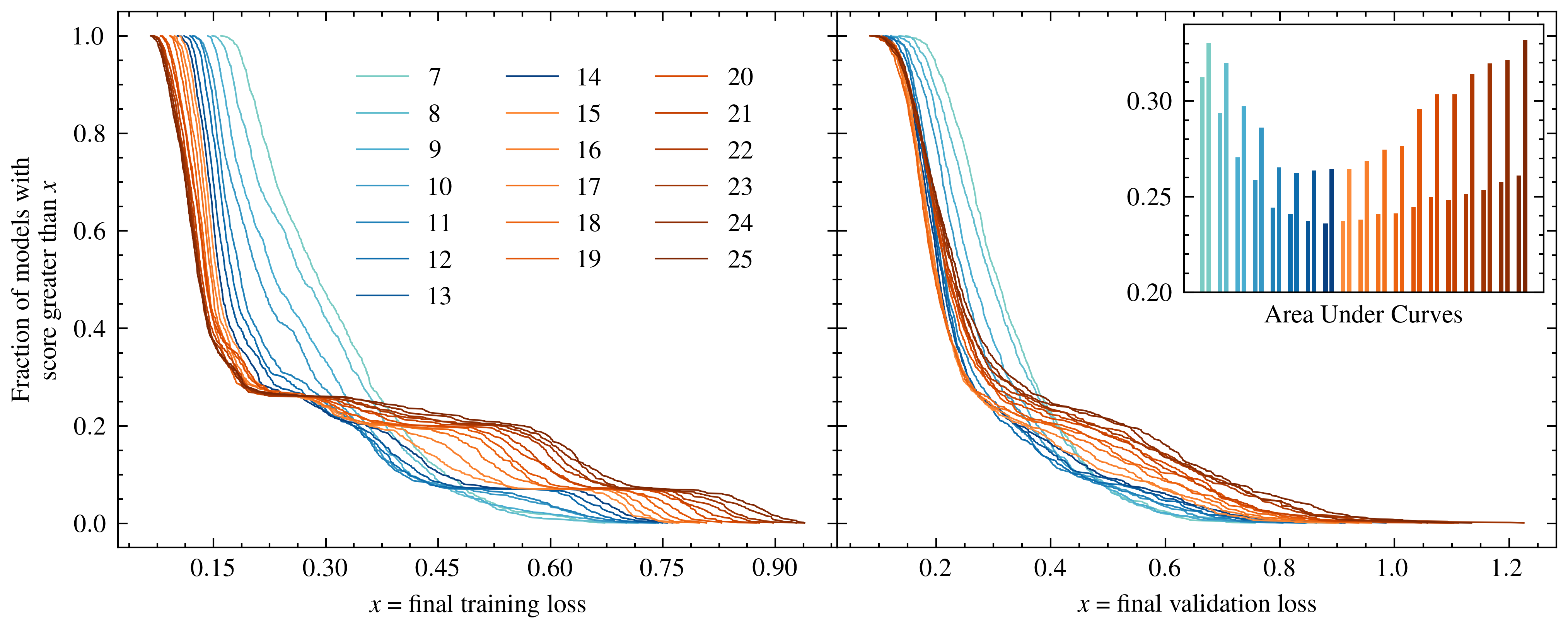}}
\caption{
    Each line is a cumulative plot of the final losses obtained by 1,000 models from the entire hyperparameter search with a fixed number of repeating layers in the quantum circuit. A point on the line at $x$ marks the fraction of runs that obtained a minimum loss of $x$. The blue gradient denotes circuits that are uncontrollable, and the red controllable. The top right plot on the right hand side denotes the area under the runs of the left and right plot, each line is grouped into pairs of bars where the left bars are the areas of the training losses, and the right the validation losses. The inflexion in the `Area Under Curves' confirms that the models best at minimising their validation and training loss over a hyperparameter search are those at the onset of controllability.
}
\label{fig:trianing-perf-profile}
\end{figure*}

\subsection{Overcontrolled circuits overfit}

The motivating aspect in using controllable circuits, and this paper's hypothesis, is testing if satiating a quantum circuit with the critical number of parameters to make it controllable provides an advantage in minimising the training and validation losses of the models over the hyperparameter search. 

An overview of the results of our experiments is presented in Fig. \ref{fig:hyp-heatmap}. This figure encapsulates the bulk training behaviour of the circuits, and begins to qualitatively address our hypothesis that controllable circuits are advantageous for minimising the validation and training loss. This is quantitatively backed by Fig. \ref{fig:trianing-perf-profile}. Moreover, it is an example of the complex behaviour introduced by hyperparameter tuning and its consequences for the trainability of quantum machine learning models.

The training losses in Fig. \ref{fig:hyp-heatmap} show that models which reached the lowest and highest losses were in their controllable regime. This is seen by noting that the lightest and darkest coloured cells are below the horizontal line respectively. Concentrating on the indexed hyperparameters with learning rate $\alpha \in \{ 0.001, 0.01\}$, these runs qualitatively support our hypothesis that controllable circuits are able to consistently produce the lowest losses. Otherwise, the other hyperparameters struggled to minimise their loss across all number of repeating layers, and strikingly obtained higher losses in their controllable regime. In this sense, the runs contradict out hypothesis, and are an example of how hyperparameters greatly influence the performance of models and their training. 

When comparing both training and validation losses of Fig. \ref{fig:hyp-heatmap}, a distinction is the horizontal light bar that appears within repeated layers 11-18 of the validation losses. This bar contrasts the training that generally continues to decrease as the number of layers increase. This is reminiscent of the fact the models are beginning to overfit: the training loss continues to be minimised whilst the validation loss increases when trainable parameters are added. This is particularly noticeable in runs when the scaling parameter is greatest. This onset of overfitting, or the centre of this light bar, is at the 15th repeated layer which is at the onset of controllability. With this information, the support of the hypothesis is strengthened as the addition of parameters past the 15th layer should not allow the model to explore any new unitary representations of the quantum circuit. Instead, in many runs the model utilises these extra parameters in other ways such as overfitting the data. Overall, controllability predicts the addition of redundant trainable parameters in increasing the expressivity of the model which allows the model to utilise these additional parameters in either overfitting, or continuing to minimise its train and validation score for a selected few hyperparameters. Indexed hyperparameter runs 1-9 inclusive are the most visually distinct examples of the latter observation.

Despite the hyperparameter $\alpha=0.1$ causing controllable circuits to become detrimental in finding minimum losses, these are in a minority. In this experiment on the modified MNIST-1D dataset, the majority of models benefited from controllability when minimising their training and validation losses.

To generate qualitative results, the cumulative sum of each row in Fig. \ref{fig:hyp-heatmap} is plotted to produce Fig. \ref{fig:trianing-perf-profile}. Similar artefacts in the training are present in both plots, in particular the three distinct columns of hyperparameter runs with a learning rate of $\alpha=0.1$ present in the training losses.

The confirmation of the overfitting is present in the right upper subplot of the validation plot of Fig. \ref{fig:trianing-perf-profile}, where the area under the curves are plotted. This is equivalent to summing the losses for each of the ten models from every hyperparameter trial with some fixed number of repeating layers. As the light bar is dominant in Fig. \ref{fig:hyp-heatmap}, the area plot shows that the total sum of losses is minimised at the onset of controllability, which might be  a computational phase transition that occurs in moving from an uncontrollable regime into a controllable regime. This implies that the hyperparameter runs which offer the greatest chance to generate models that minimise their validation loss is again at the onset of controllability, where circuits with less or a greater number of parameters fail to minimise their training loss or overfit the data respectively.

Additionally, Fig. \ref{fig:trianing-perf-profile} reinforces the fact that adding additional parameters to controllable circuits will obtain the best and worst performing models overall, in effect becoming a lottery pick of finding a circuit with hyperparameter tuning that learns the task best. 

Overall, the results in Fig. \ref{fig:hyp-heatmap} and Fig. \ref{fig:trianing-perf-profile} confirm the hypothesis in the case of this learning task: controllable circuits increase the success rate in finding circuits that minimise their training and validation loss. However, the hypothesis is made sharper in these results where circuits just at the onset of controllability are most robust in minimising their training and validation loss with respect to changes in hyperparameters.

\subsection{Near controllable circuits predict later winners}

As hyperparameter searches are costly, it is always advantageous to find reasons to reduce the search space. 

\begin{figure}[hbtp]
\centerline{\scalebox{0.65}{
\includegraphics{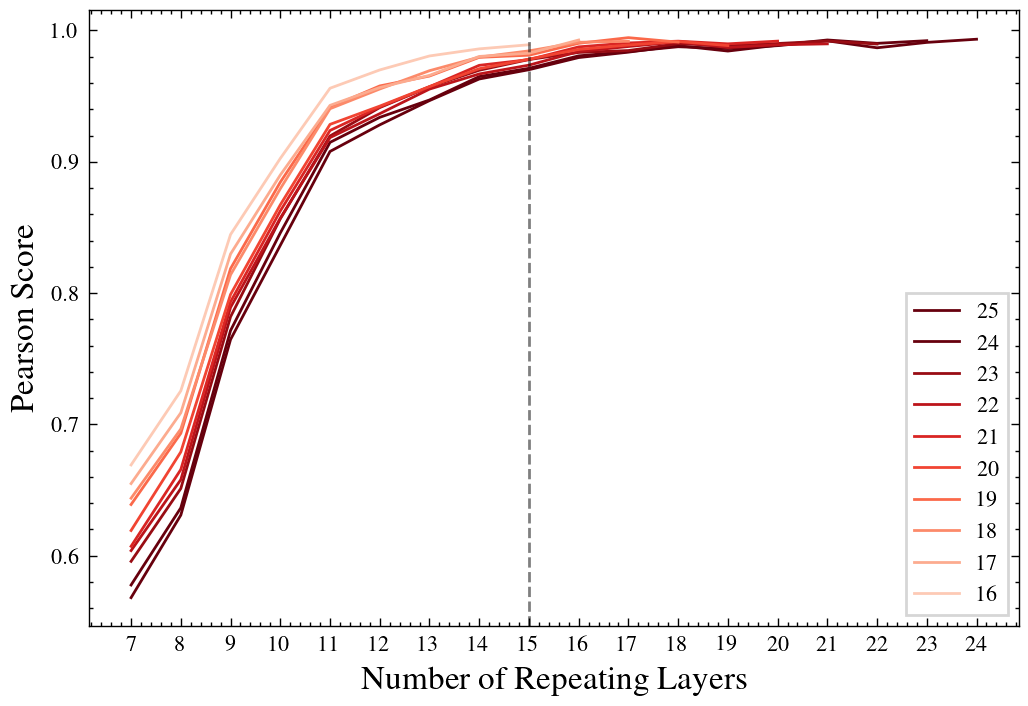}
}}
\caption{Each line, labelled $r$, corresponds to a hyperparameter search of 100 trials with circuits of $r$ number of repeated layers. The value for integer $x$ along the x-axis is the Pearson correlation coefficient computed across the aggregated final validation losses obtained by the 100 models with $r$ repeating layers, against the same hyperparameter search of circuits containing $x$ number of repeated layers. For every Pearson score the p-tests were $<0.001$. Circuits with 15 or more repeating layers are controllable, and so the validation loss values obtained by such circuits have a strong linear correlation with the losses achieved by circuits that are near the onset of controllability.}
\label{fig:validation_corr}
\end{figure}

Fig. \ref{fig:validation_corr} demonstrates that the final validation losses obtained for controllable circuits have a strong linear correlation with the final validation losses of circuits with a greater number of repeating layers. Notably, much weaker linear correlations are found for circuits with 10 repeating layers or less, making prediction hard in estimating circuits that will obtain minimal validation losses when increasing the number of repeated layers. Apart from the 7th repeated layer, the Pearsons scores are approximately monotone increasing with the number of repeating layers. At the point of the circuits becoming controllable, the Pearsons score is greater than 0.96, showing that the best performing circuits with 15 repeated layers have a high linear correlation with the best performing circuits that have additional layers.

These results hint at a possible technique for reducing the computational costs of hyperparameter studies while still obtaining the best performing circuits. First, one performs the hyperparameter search fixed at the onset of controllability, 15 layers in our experiments, then restrict the hyperparameter search to the space spanned by the most performant circuits with 15 layers, and then add additional layers to these selected models before further training. Further experiments would need to be conducted to verify this conjecture.

\section{Conclusion} \label{sec:conclusion}

Introducing controllable circuits for variational hybrid quantum models theoretically maximises the freedom for the machine to find the global minimum. In essence, it allows the machine to represent any operator in the circuit's Hilbert space, or in other words, any quantum circuit within the qubit space. Therefore controllable circuits maximise the expressivity that a unitary circuit ansatz can offer. 

This work empirically demonstrated for a subset of MNIST-1D, that the majority of models leveraged their controllability to increase the success rate in minimising both their validation and training loss over a hyperparameter sweep. 

Moreover, the results support the theory of controllable circuits maximising their expressivity as the majority of controllable models with additional parameters were overfitting their data; the extra parameters were redundant in expressing new quantum gates and so the quantum machine learning model utilised them elsewhere. 

These benefits can be further exploited to reduce the range of the hyperparameter searches, as there was a strong correlation existing between the best performing circuits that were at the onset of controllability, and the best circuits that were saturated with additional parameters. Therefore the best performing circuits that are just over the threshold of controllabiltiy can be selected for further training, with a strong likelihood of obtaining circuits that continue to reduce their validation and training loss when increasing their parameter count.

Lastly, the HEA ansatz that was uncontrollable for all repeated layers mimicked the behaviour of the controllable ansatz. The only similarity was the encoding that occupied the smaller subspace. This suggests that the model was indifferent to training a circuit with more generators than what were used to generate the encoding operator, and thus highlights the influence of the embedding space defined by the encoding on the training process of the model. 

There are many questions that arise after analysing these results. For example, two circuits can be controllable on the same space, yet one may utilise a different ansatz which results in the circuit parameters to be distributed differently amongst the generators in the DLA. Despite the circuits having the same expressivity, the trainability may change due to the different architectures of the quantum model. Testing this would be akin to introducing new ansatzes to the experiments which have controllability on the same subspace and equal DLA.

Furthermore, the observation of overfitting in a large majority of models hints at the possibility to use regularisation methods in order for the model to increase its generalization abilities. This will be further investigated in future experiments through analysing the training-validation gap and effects of regularisation in the training.

Finally, the number of classical datasets for the learning task needs to be increased. This would inevitably introduce more unknowns, perhaps the largest being the type of preprocessing and encoding, however given that the dataset used in this work is already a multi-classification problem with noise, the results of this paper should continue to hold in more complicated cases. Additional studies are planned to confirm this hypothesis.



\end{document}